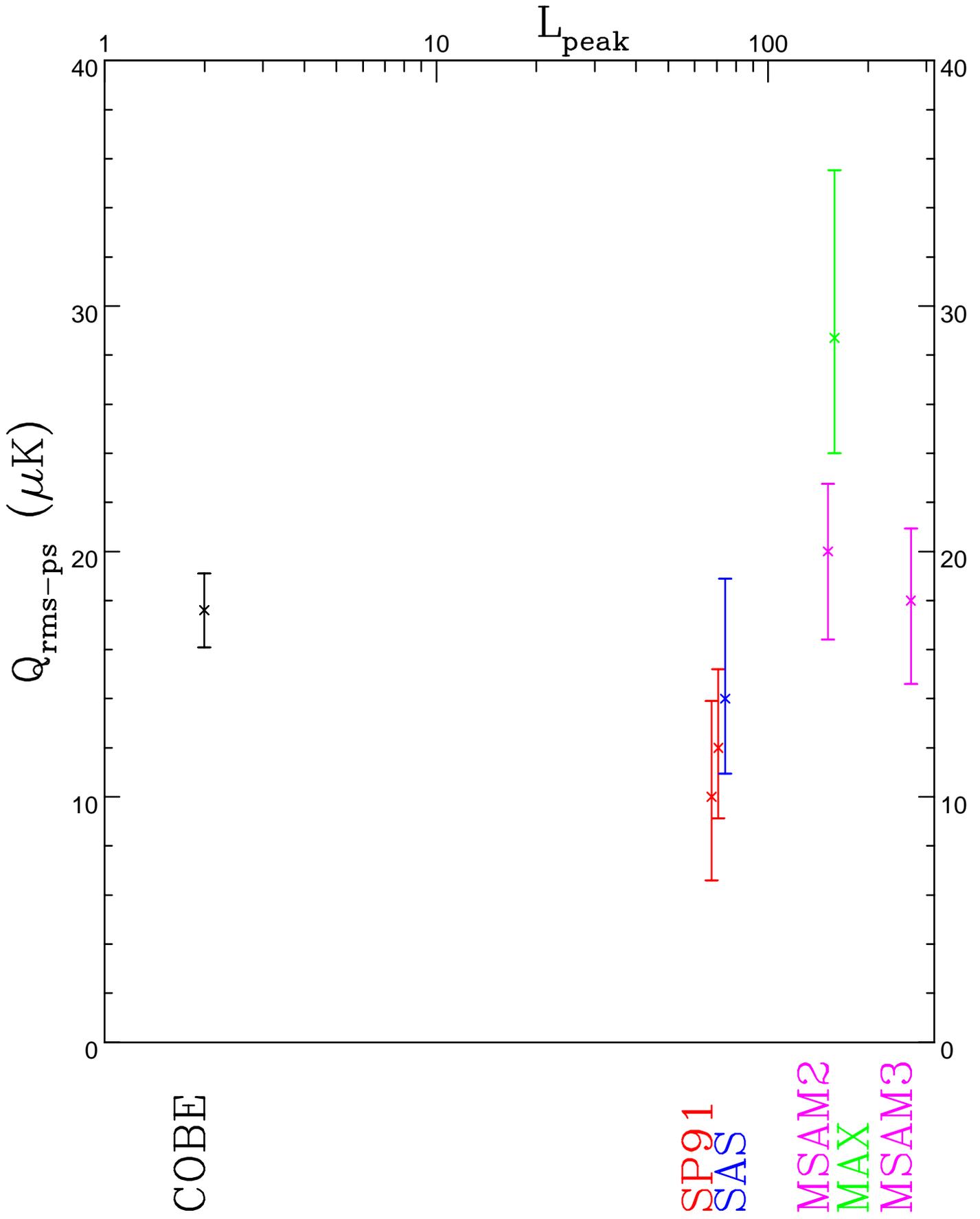

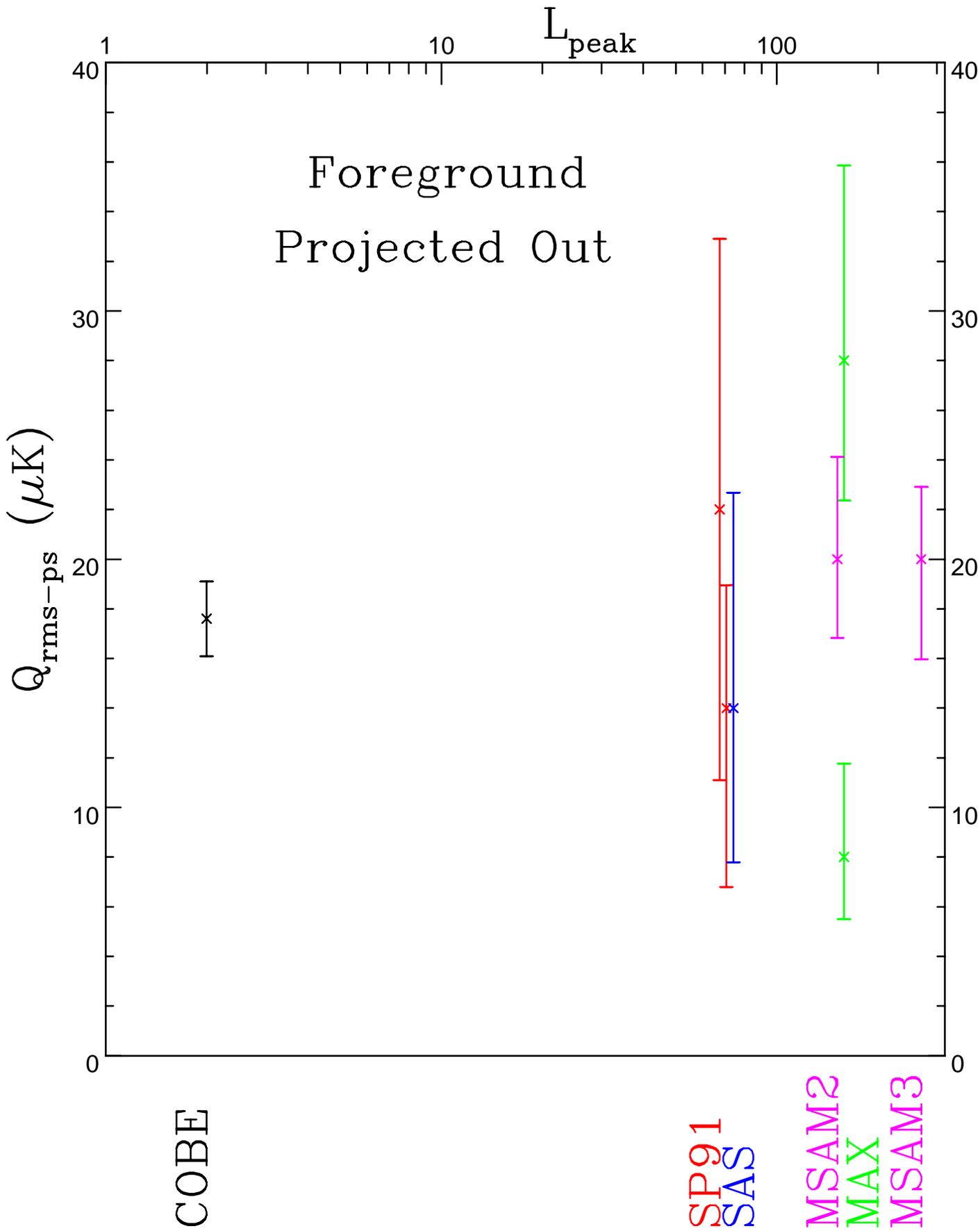

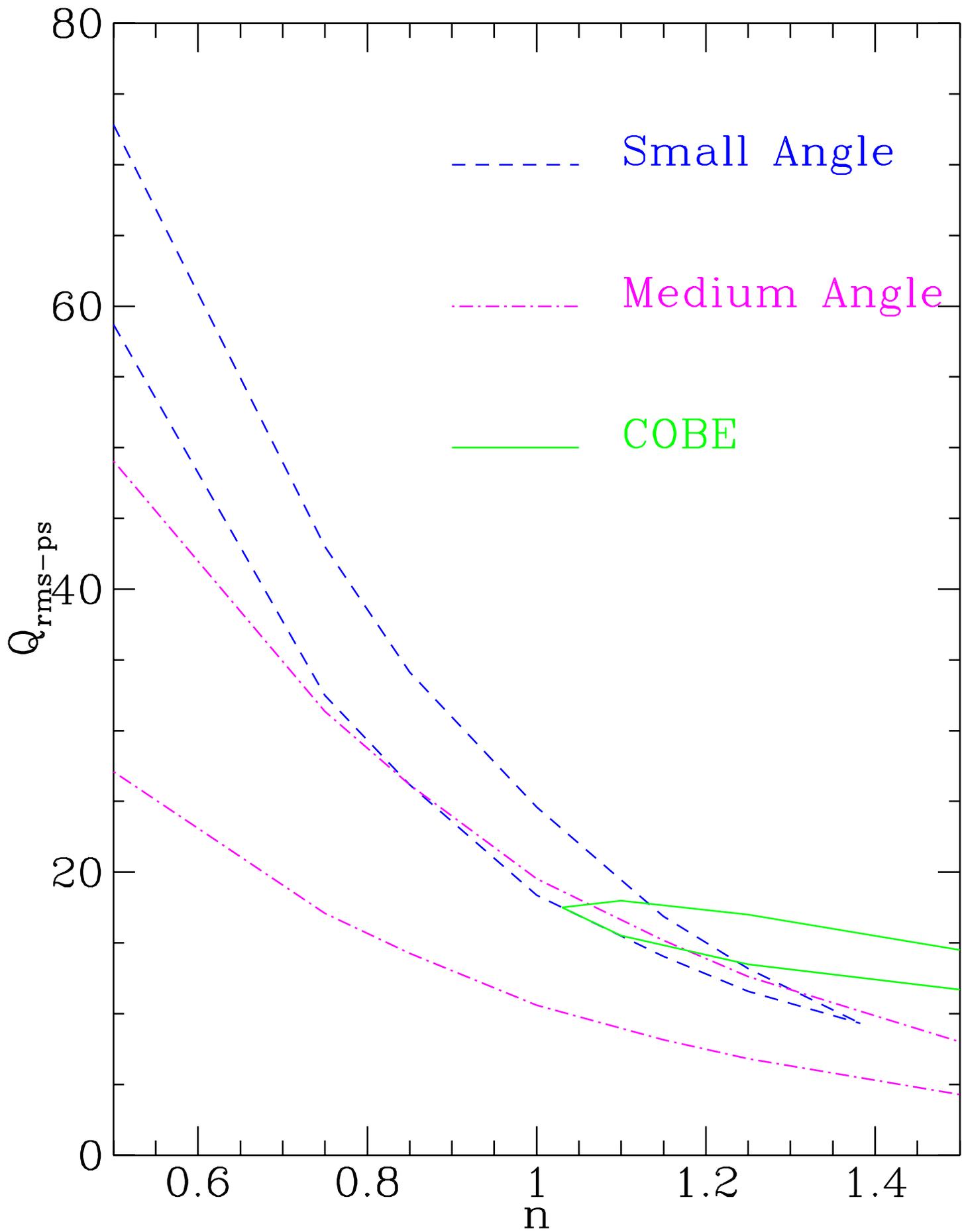

# Analysis of Small and Medium-Scale Cosmic Microwave Background Experiments

Scott Dodelson[*]

*NASA/Fermilab Astrophysics Center, Fermi National Accelerator Laboratory, Batavia, IL 60510-0500*

Arthur Kosowsky[†]

*Department of Physics, Enrico Fermi Institute, The University of Chicago, Chicago, IL 60637-1433.*

(October, 1994)

## Abstract

Anisotropies in the temperature of the cosmic microwave background have been detected on a range of scales by several different experiments. These anisotropies reflect the primordial spectrum of metric perturbations in the early universe. In principle, the largest barrier to a clean interpretation of the experimental results is contamination by foreground sources. We address this issue by projecting out likely sources of foreground contamination from seven separate small-angle and medium-angle experiments. We then calculate likeli-hood functions for models with adiabatic perturbations, first for the amplitude of the spectrum while constraining the spectral index to be $n = 1$, and then jointly for the amplitude and spectrum of the fluctuations. All of the experiments are so far consistent with the simplest inflationary models; for $n = 1$ the experiments' combined best-fit quadrupole amplitude is $Q_{\mathrm{rms-ps}} = 18^{+3}_{-1} \, \mu K$, in excellent agreement with the COBE two-year data. In ($Q_{\mathrm{rms-ps}}$, n) space, the allowed region incorporating intermediate and small-scale experiments is substantially more constrained than from COBE alone. We briefly discuss the expected improvement in the data in the near future and corresponding constraints on cosmological models.

---

[*]dodelson@virgo.fnal.gov

[†]akosowsky@cfa.harvard.edu. Current address: Harvard-Smithsonian Center for Astrophysics, Mail Stop 51, 60 Garden St., Cambridge, MA 02138.

98.70.Vc

More than any other cosmological observation, measurements of temperature anisotropies in the Cosmic Microwave Background (CMB) are a direct probe of the primordial spectrum of metric perturbations. Precise measurements will give both the amplitude and spectral index of the perturbation spectrum, may allow disentangling scalar from tensor perturbations, and will provide information on the origin of the perturbations (whether from inflation or some type of topological defect, for example).

Measurements of CMB temperature anisotropies are notoriously difficult and have only recently attained the necessary signal-to-noise ratio for meaningful results. Since the COBE satellite made the first anisotropy detection in 1992 [1], nearly a dozen other experiments have announced positive detections on a wide range of angular scales at amplitudes of a few parts in $10^5$ of the background temperature [2]. At such small temperature differences, the main experimental obstacle is instrumental noise; technological advances and experimental ingenuity have pushed noise levels down by a factor of $10^4$ in the past 30 years, and this trend will continue for the coming few rounds of experiments. The largest hurdle in principle is disentangling the foreground contribution: for a given measurement, how much of the signal comes from the blackbody CMB and how much from other sources of microwave radiation?

Two different techniques are useful for sorting out the foreground. The first is to extrapolate sky maps at other frequencies (e.g. radio maps) to estimate the microwave emission in various parts of the sky, and then subtract this from the measured signal to obtain the cosmic signal. This process depends on detailed modeling of various sources and involves uncontrolled extrapolations over large frequency ranges. In this paper we focus on the complementary method of using measurements at multiple frequency channels to subtract out the non-blackbody piece of the measured signal. Clearly, as measurements improve, a combination of the two methods will give the most reliable interpretation of experiments; here we show how much various measurements may be affected by foreground contamination.

We analyze the current small-scale and medium-scale anisotropy measurements which employ multiple frequencies and for which data is publicly available: MAX [3], MSAM [4], South Pole [5], and Saskatoon [6]. The first two are balloon-born packages, while the other two are based on the ground. All measure in either three or four frequency channels. Saskatoon measures at the largest angular scale, roughly two degrees, and South Pole is slightly smaller; we call these two experiments "medium-angle." MSAM and MAX measure at smaller scales, roughly 20 to 30 arc-minute scales; these are referred to as "small-angle." The angular scale is more precisely characterized by the window function $W_\ell$ of each experiment, defined as follows: If the sky temperature with mean $T_0 = 2.735°$ K is decomposed into spherical harmonics as $T = T_0(1 + \sum_{\ell m} a_{\ell m} Y_{\ell m})$, then the experiment will measure a mean temperature variance (in the absence of noise) given by

$$\left(\frac{\delta T}{T_0}\right)^2 = \sum_{\ell=2}^{\infty} \frac{W_\ell}{4\pi} \sum_{m=-l}^{l} |a_{lm}|^2 \,. \tag{1}$$

For the purposes of the likelihood analysis presented below, the window function is extended to covariances between temperatures measured at different points on the sky. Each experiment has a unique window function, which incorporates the beam profile, chopping strategy, scan pattern on the sky, and other experimental details. We have calculated window functions for the experiments considered here to an accuracy of around 1%.

We employ the well-known method of Bayesian likelihood analysis [7] to the data from each of these experiments. Likelihood analysis asks how likely is a given set of measurements if a particular theory is true. The likelihood function quantifying this probability is

$$\mathcal{L} = \frac{(2\pi)^{-N/2}}{\sqrt{\det C}} \exp\left[-\frac{1}{2} D^T C^{-1} D\right]. \tag{2}$$

Here $D$ is a data vector of length $N$; in this case, $D$ contains values of the temperature anisotropy measured at $N_p$ patches on the sky with $N_c$ different channels (either at different frequencies or polarization states), such that $N_p N_c = N$. The correlation matrix $C$ is the expectation value $\langle DD^T \rangle$ which has two separate pieces, the instrumental noise and the theoretical signal. The latter depends on the parameters in the theory being tested. Given a likelihood function depending on a two-parameter $(p_1, p_2)$ theory, we obtain an allowed $1\sigma$ region in $(p_1, p_2)$ by the condition

$$\int_\Gamma dp_1 \, dp_2 \, \mathcal{L}(p_1, p_2) = 0.68; \qquad \mathcal{L}(\Gamma) \quad \text{constant}, \tag{3}$$

where the boundary of the allowed region is denoted by $\Gamma$. For a single parameter, the region reduces to an interval and the boundary is its endpoints.

The above analysis is standard practice. Before applying this analysis to the data, we use the multiple frequency channels available to discriminate against foreground contamination [8]. In particular, for each experiment we choose one source of foreground judged most likely to be a contaminant: for MAX and MSAM dust emission, while for South Pole and Saskatoon synchrotron emission. These choices are based on the frequency ranges of the experiments: the high frequency experiments are more likely to be sensitive to dust while the low frequency ones more sensitive to free-free and syncrotron emission. This component is then assigned a given spectral index. For example, the signal due to free-free emission is assumed to scale with frequency as $S(\nu) = S(\nu_1)(\nu_1/\nu)^{2.1}$. If a given experiment has only two frequency channels, then the linear combination

$$\tilde{D} \equiv D(\nu_1) - (\nu_2/\nu_1)^{2.1} D(\nu_2) \tag{4}$$

is completely independent of free-free contamination. Note that if the frequencies are closely spaced, then $\tilde{D}$ approaches the difference between the data in the two channels. This difference is zero for a cosmic signal (expected to be frequency independent), so the signal to noise ratio becomes very small in the limit of closely spaced frequencies. All other factors being equal, experiments that cover a large range of frequencies are best able to distinguish cosmic signal from foreground contaminants. This advantage shows up noticably in our analysis. The extension of this method to more channels is straightforward: we always choose the $N_{\text{channel}} - 1$ linear combinations of the data that are independent of foreground contamination from the assumed source. Conversely, an $N$-channel measurement can in principle be used to project out $N - 1$ different foreground sources, but in practice the signal-to-noise ratios are small enough that substantial redundancy is necessary for reasonably significant results.

The likelihood and foreground analysis described above is completely general. We perform the Bayesian likelihood analysis on the class of theories, based on inflation, with a primordial gaussian spectrum of density perturbations

$$\langle (\delta\rho/\rho)^2 \rangle \propto k^n$$

with $k$ the wavenumber of the perturbation. First we perform the analysis with the spectral index fixed at $n = 1$ and the amplitude $Q_{\mathrm{rms-ps}}$ [9] as the only free parameter of the theory, and then we consider a two-parameter theory depending on both $n$ and $Q_{\mathrm{rms-ps}}$.

This two parameter inflationary theory is well-developed and detailed predictions of CMB temperature anisotropies are readily available [10]. The actual CMB temperature anisotropies depend also on a variety of cosmological parameters: the Hubble constant $H_0 = 100h\,\mathrm{km}\,\sec^{-1}\mathrm{Mpc}^{-1}$, the baryon mass fraction $\Omega_b$, the nature of the dark matter, the equivalent mass fraction in a cosmological constant $\Omega_\Lambda$, the tensor perturbation spectrum, and the redshift of reionization $z_R$. All of these parameters may vary in inflationary models [11]. Nonetheless, recent work [12] has demonstrated that to within 10% the CMB anisotropies depend only on an effective spectral index defined by

$$\tilde{n} \equiv n - 0.28 \ln(1.56 - 0.56n) - 0.00036 z_R^{3/2} + 0.26 \left( 1 - 2h\sqrt{1 - \Omega_\Lambda} \right).\tag{5}$$

This relationship assumes the preferred nucleosynthesis relation $\Omega_b h^2 = 0.0125$ [13] and the tensor spectrum conditions $n_T = n - 1$ and $r \equiv C_2^{(T)}/C_2 \approx 7(1 - n)$ given by the simplest inflation models [14]. The effects of relaxing these conditions has been explored elsewhere [15]; it appears that even if these conditions are relaxed the theory can still be parameterized in terms of its amplitude and a generalized form of Eq. (5). Thus shifting the value of these cosmological parameters only moves to a different place in ($\tilde{n}$, $Q_{\mathrm{rms-ps}}$) space. This "cosmic confusion," as it has been dubbed [12], means that the CMB will never be a strong positive test for any particular set of cosmological parameters. This pessimistic outlook was the conclusion of Ref. [12] and has since been often reiterated. On the other hand, we want to emphasize that this same effect makes the CMB an extremely powerful negative test: if various experiments end up being inconsistent for a particular set of parameters, Eq. (5) shows that they rule out all other inflationary scenarios which involve only shifting cosmological parameters.

Proceeding with the experiment analysis, we first fix $\tilde{n} = 1$ and consider the conditional likelihood for the perturbation amplitude given by the small and medium-angle experiments. The best fit value and $1\sigma$ range for $Q_{\rm rms-ps}$ are plotted for each of seven data sets in Fig. 1. The $x$-axis shows the region in $\ell$-space to which each experiment is most sensitive; the angular scale in degrees is given roughly by $100/\ell$. COBE (which we have not analyzed) is sensitive to large angular scales and is displayed for reference at a small value of $\ell$ [16].

Figure 1a plots the allowed range of $Q_{\rm rms-ps}$ neglecting the effect of foreground; Fig. 1b projects out foreground contamination as described above. For MSAM and MAX, projecting out foreground makes little difference in the error bars because these experiments cover a large range of frequencies [17]. For Saskatoon and South Pole the error bars on $Q_{\rm rms-ps}$ become much larger when foreground is projected out [8]; note especially the South Pole experiments which cover the narrow frequency range $25 - 35$ GHz. This is because the effective signal to noise ratio is small after projecting out background from a signal spanning a narrow frequency range, as noted above. In every case, the $\chi^2$ per degree of freedom for the most likely amplitude is reduced by projecting out foreground. This suggests that fitting just one blackbody component to the measurements is not sufficient: projecting out at least one foreground component is essential.

The agreement between large and small scale observations for the $\tilde{n} = 1$ model is remarkably good. The combined best fit from the seven medium/small scale experiments is $Q_{\rm rms-ps} = 18^{+3}_{-1}\mu$K, in complete agreement with the COBE two-year values of $20.4 \pm 1.7\mu$K [16] and $17.6 \pm 1.5\mu$K [18] . The uncertainty on the COBE measurement is not going to get much smaller, being dominated by cosmic variance. Cosmic variance is not yet a major factor for the smaller scale experiments; the uncertainty on $Q_{\rm rms-ps}$ from these experiments is essentially limited only by how well the foreground contamination can be eliminated.

For a two-parameter fit, we now allow $\tilde{n}$ to vary and find the allowed region in ($Q_{\rm rms-ps}$, $\tilde{n}$) space for the small and medium-angle experiments. At large angles, the COBE team has performed a similar analysis; we quote their results here [18]. Figure 2 shows the regions allowed at a $1\sigma$ level by COBE, the medium angle experiments and the small angle experiments after projecting out one foreground component. A large region of consistency is currently allowed. Clearly at this stage of the observations, the medium and small-angle experiments have uncertainties which are too large to make this a powerful test. The medium angle experiments in particular have large uncertainties due to the lack of frequency coverage. In the near future the situation will improve [2]; in fact, both Saskatoon and South Pole have recently been redone with larger frequency coverage, which should substantially reduce the error bars. If the allowed regions for the three classes of experiments eventually do not overlap, this will be strong evidence against the inflationary scenario.

Another noticeable feature of Fig. 2 is the different slopes of the allowed regions for the three different types of experiments. The large angle results are least sloped since COBE is sensitive to the lowest order multipoles, those closest to the normalization point at $\ell = 2$. A tilt in the spectrum thus has a relatively small effect on $Q_{\mathrm{rms-ps}}$. The small angle experiments are most affected by a spectral tilt, with their best fit $Q_{\mathrm{rms-ps}}$ being significantly reduced (increased) for $n$ large (small). The variations in slope are essential characteristics of experiments on differing angular scales, ultimately leading to a powerful test of inflationary models.

To conclude, we have analyzed small and medium scale anisotropy experiments by projecting out one foreground component. For experiments with wide frequency coverage, this procedure does not substantially increase the error bars on the parameters in a theory. All current medium and small-angle experiments with multiple frequency channels are consistent with simple inflation models normalized to COBE, with error bars comparable to COBE. The inflation model predictions are degenerate in the cosmological parameters, making the theory of inflation testable since only two free parameters determine the anisotropies. The best way to test this theory at present is through CMB experiments at different angular scales. Current medium angle data lack the frequency coverage to discriminate effectively against foreground, so the test is not yet very powerful. However, in the next three to five years, we expect improving signal-to-noise ratios and wider frequency coverage will test inflationary models in $(Q_{\mathrm{rms-ps}}, \tilde{n})$ space at the 10% level.

## ACKNOWLEDGMENTS

We thank Stephan Meyer and Lyman Page for patient explanations of their experiments. This work was supported in part by the DOE (at Chicago and Fermilab) and by NASA through grant No. NAGW-2381 (at Fermilab). AK was supported in part by the NASA Graduate Student Researchers Program.

# REFERENCES

[1] G.F. Smoot *et al.*, Ap. J., **396**, L1 (1992).

[2] For a recent review of some current experimental results and associated difficulties see M. White, D. Scott, and J. Silk, Ann. Rev. Astro. Astrophys. (in press); L. Page, in Proceedings of Workshop on Cosmic Background Radiation Two Years After COBE, ed. L. Krauss (World Scientific, in press 1994).

[3] This data is from two scans of MAX3: J. Gundersen *et al.*, Ap. J. **413**, L1 (1993); P. Meinhold *et al.*, Ap. J. **409**, L1 (1993).

[4] E. S. Cheng *et al.*, Ap. J. **422**, L37 (1994) used two independent chopping schemes (MSAM2 and MSAM3) to look at the same sky at two different angular scales. Their analysis (presented in the "raw" graph in Figure 1) includes the effects of dust in a slightly different way than we do. We utilize the entire MSAM data set; that is, we do not drop the regions of data which possibly contain unresolved sources. Hopefully further study of the same region will soon resolve whether sources are actually present.

[5] We analyze two scans taken at the South Pole in 1991: in T. Gaier *et al.*, Ap. J. **398**, L1 (1992); J. Schuster *et al.*, Ap. J. **412**, L47 (1993).

[6] Data taken in Saskatoon in 1993: E. J. Wollack *et al.*, Ap. J. **419**, L49 (1993).

[7] A. C. S. Readhead *et al.*, Ap. J. **346**, 566 (1989).

[8] S. Dodelson and A. Stebbins, Ap. J. **433**, 440 (1994).

[9] In a given model, $Q_{\mathrm{rms-ps}}$ can be related to the amplitude of the initial power spectrum, $A \equiv P(k)/k^n$. For example, if anisotropies are generated only by scalar perturbations to the metric, then

$$A = (Q_{\mathrm{rms-ps}}/T_0)^2 \frac{64\pi}{5H_0^{n+3}} \frac{\Gamma\left(\frac{9-n}{2}\right)\Gamma^2\left(\frac{4-n}{2}\right)}{\Gamma\left(\frac{3+n}{2}\right)\Gamma\left(3-n\right)}.$$

See P.J.E. Peebles, *The Large Scale Structure of the Universe* (Princeton: Princeton University Press, 1980).

[10] P. J. E. Peebles and J. T. Yu, Ap. J. **162**, 815 (1970); M. L. Wilson and J. Silk, Ap. J. **243**, 14 (1981); J. R. Bond and G. Efstathiou, Ap. J. **285**, L45 (1984); N. Vittorio and J. Silk, Ap. J. **285**, 39 (1984); N. Sugiyama and N. Gouda, Prog. Theor. Phys. **88**, 803 (1992); S. Dodelson and J.M. Jubas, Phys. Rev. Lett. **70**, 2224 (1993); R. Crittenden, *et al.*, Phys. Rev. Lett. **71**, 324 (1993).

[11] We assume $\Omega = 1$ for this analysis, as this is a generic prediction of inflation. Open universes have been analyzed by M. L. Wilson and J. Silk, Ref. [10]; L.F. Abbot and R.K. Schaefer, Ap. J. **308**, 546 (1986); M. Kamionkowski and D.N. Spergel, submitted to Ap. J. (1994); M. Kamionkowski, D.N. Spergel, and N. Sugiyama, submitted to Ap. J. Letters (1994).

[12] J.R. Bond *et al.*, Phys. Rev. Lett. **72**, 13 (1994)

[13] T.P. Walker *et al.*, Ap. J. **376**, 51 (1991).

[14] R.L. Davis *et al.*, Phys. Rev. Lett. **69**, 1856 (1992).

[15] S. Dodelson, L. Knox, and E.W. Kolb, Phys. Rev. Lett. **72**, 3444 (1994) studied the effect of relaxing the conditions on the tensor perturbation spectrum, allowing the amplitude of the tensor modes to be another free parameter. Previous work varying $\Omega_B$ and $h$ was carried out by J.R. Bond, G. Efstathiou, P. M. Lubin, and P.R. Meinhold, Phys. Rev. Lett. **66**, 2179 (1991); S. Dodelson and J.M. Jubas, Ref. [10]; K. Gorski, R. Stompor, and R. Juszkiewicz, Ap. J. **410**, L1 (1993).

[16] K.M. Górski *et al.*, Ap. J. in press (1994).

[17] In Fig. 1b, points from both the MAX GUM scan and the $\mu$PEG scan are plotted. The $\mu$PEG scan is apparently highly contaminated by dust emission and is off the scale of Fig. 1a displaying the uncorrected data. Subsequent MAX scans in different regions of

the sky have confirmed the result from the GUM scan.

[18] C. L. Bennett *et al.*, Ap. J., in press (1994).

FIG. 1. The results of seven medium and small-angular scale CMB experiments, plotted as the amplitude $Q_{\mathrm{rms-ps}}$ versus angular scale in multipole number. The effective spectral index is assumed to be $\bar{n} = 1$. The error bars represent $1\sigma$ deviations. Figure 1a shows the raw measurements; Fig. 1b shows the same measurements after a likely foreground contaminant has been projected out.

FIG. 2. Likelihood contours plotted in $(Q_{\mathrm{rms-ps}}, \bar{n})$ space. The overlap region between small, medium, and large-angle experiments is currently allowed at a 68% confidence level.